\pdfoutput=1

\documentclass[twocolumn]{aastex63}
\usepackage[normalem]{ulem}
\usepackage{bm}
\usepackage[utf8]{inputenc}
\usepackage{amssymb}
\usepackage{amsthm}
\usepackage{amsmath}
\usepackage{mathrsfs}

\begin{document} 
\title{How to search for multiple messengers - a general framework beyond two messengers} 
\correspondingauthor{Do\u{g}a Veske}
\email{dv2397@columbia.edu}

\author[0000-0003-4225-0895]{Do\u{g}a Veske}
\affiliation{Department of Physics, Columbia University in the City of New York, New York, NY 10027, USA}

\author[0000-0003-1306-5260]{Zsuzsa M\'arka}
\affiliation{Columbia Astrophysics Laboratory, Columbia University in the City of New York, New York, NY 10027, USA}

\author[0000-0001-5607-3637]{Imre Bartos}
\affiliation{Department of Physics, University of Florida, PO Box 118440, Gainesville, FL 32611-8440, USA}

\author[0000-0002-3957-1324]{Szabolcs M\'arka}
\affiliation{Department of Physics, Columbia University in the City of New York, New York, NY 10027, USA}

\begin{abstract}
Quantification of the significance of a candidate multi-messenger detection of cosmic events is an emerging need in the astrophysics and astronomy communities. In this paper we show that a model-independent optimal search does not exist, and we present a general Bayesian method for the optimal model-dependent search, which is scalable to any number and any kind of messengers, and applicable to any model. In the end, we demonstrate it through an example for a joint gravitational wave, high-energy neutrino, short gamma-ray burst event search; which has not been examined heretofore.
\end{abstract}

\section{Introduction}
\label{sec:Introduction}
Astronomy has started via observations made in the visible region of the electromagnetic spectrum in ancient times \citep{gobekli,hoskin1999cambridge}. As the technology and physics knowledge of humanity developed, more and better observations were made with new equipment and via new messengers; such as the whole electromagnetic spectrum \citep{radio,RevModPhys.75.995,Opal702,ir,1965ApJ...142..419P,1990RMxAA..21..459F}, cosmic rays \citep{hess,Sommers_2009}, neutrinos \citep{PhysRevLett.58.1490,PhysRevLett.20.1205} and recently gravitational waves \citep{PhysRevLett.116.061102}. The new messengers have made it possible to observe events which had not been possible before, as well as to gather a more complete picture of a single event by probing different processes of it. This allows us to understand the ongoing physics at extreme conditions that we cannot produce on Earth.

Three observations, each involving at least two messengers, can be given as examples for multi-messenger discoveries. The first one was the supernova SN 1987A observed in electromagnetic waves and low-energy neutrinos (in MeV energy range) in 1987 \citep{doi:10.1146/annurev.aa.27.090189.003213}. The second was the observation of the binary neutron star merger, GW170817, which was discovered with gravitational waves and gamma-rays \citep{Abbott_2017}. Later it was tracked in all of the electromagnetic spectrum. Finally, the last one was a flaring blazar observed in gamma rays and high-energy neutrinos with $3\sigma$ significance \citep{blazar}.

As detectors improve for all messengers, it is natural to expect to have more multi-messenger detections with more messengers and better data. Therefore a need for a framework for multi-messenger coincidence quantification is inevitable. For example, the HAWC observatory recently observed a subthreshold gamma-ray candidate coming from the coincident sky area of a neutrino detected by IceCube in response to a significant gravitational wave detection by LIGO and Virgo detectors \citep{hawc}.

One challenge here is relating different messengers of the same source to each other. The possibility of having several unrelated detections or noise triggers coincidentally showing up in the appropriate spatial and temporal regions for a potential multi-messenger observation makes it impossible to deduce the multi-messenger detection with absolute certainty. Therefore a statistical inference has to be made \citep{PhysRevD.100.083017,2008CQGra..25k4039A,PhysRevD.85.103004,raven,Ashton_2018}.

In this paper, we first describe the main challenge for a multi-messenger search in Sec. \ref{sec:problem} and show that a model-independent optimal search does not exist. We provide a Bayesian solution for assigning a significance to a multi-messenger detection, or candidate observations of different messengers in Sec. \ref{sec:bayesian}, which is the extension of the method described in \cite{PhysRevD.100.083017} for coincident high-energy neutrinos and gravitational waves.  In Sec. \ref{sec:Use_Cases} we demonstrate the method for a joint gravitational wave, high-energy neutrino, short gamma-ray burst event search, which has not been examined until this work. We note that the described method is scalable to any number or any type of messengers. We conclude in Sec. \ref{sec:conclusion}.

\section{The Multi-messenger Search Problem }
\label{sec:problem}
The problem we want to address in this paper is to construct an optimal search for multi-messenger events. These searches can be described as the analyses that quantify the chance of a number of messengers coming from the same source. As that number can be at least two (i.e., what is the chance that at least two of the messengers have come from the same source?), one can look for multi-messenger events with a different number of messengers, and can put a constraint to the type of the messengers as well.

In terms of statistics, the problem for these searches is a composite hypotheses testing problem. Our input parameters are the detection properties of the messengers, which may or may not be of astrophysical origin. Correspondingly, let's consider a search with $n$ $(n \geq 2)$ messengers. There are several discrete sub-hypotheses which represent $m$ ($0\leq m\leq n$) of the messengers being astrophysical and coming from $l$ ($1\leq l \leq m$) existing astrophysical sources. Naturally, all of them being noise originated $(m=l=0)$ is also a possibility. The total number of sub-hypotheses for $n$ messengers is given by $f(n+1)$, for the $f$ function defined recursively in Eq. \eqref{f}.
\begin{equation}
    f(n+1)=\sum_{i=0}^{n} \binom{n}{i}f(i),\ f(0)=1
    \label{f}
\end{equation}
For example, for two messengers, there are $f(3)=5$ sub-hypotheses, which are; both of them being not real (noise), only the first one being real, only the second one being real, both of them being real and coming from the same source, and finally both of them being real and coming from different sources.

In the context of the multi-messenger search, the possible sub-hypotheses form two distinct hypotheses, commonly named \emph{null} and \emph{alternative} hypotheses. We will call our alternative hypothesis as the \emph{signal} hypothesis. For a multi-messenger search for at least two messengers coming from the same source, the null hypothesis consists of the sub-hypotheses which have $l=m$, so that none of the astrophysical messengers have come from the same source. The signal hypothesis contains the remaining sub-hypotheses. As there is a composite hypotheses testing problem, one may naturally look for the uniformly most powerful (UMP) test, which does not exist for our problem in general as it will be illustrated. The UMP test is the level $\alpha$ test (false alarm or type I error probability for all null sub-hypotheses is at most $\alpha$) which has the highest statistical power (the least false dismissal or type II error probability) for all of the signal sub-hypotheses. If we show that the level $\alpha$ most powerful tests for any two signal sub-hypotheses are different, then we will have proved that the UMP test does not exist. It should be noted that search for UMP tests is meaningful when we have more than two signal sub-hypotheses, as for a single signal sub-hypothesis one can always find the most powerful test. Hence we construct our illustration for $n>2$, for having more than one signal sub-hypothesis. Consider the search for at least two messengers from the same source with $n=3$ with messengers: $M_1$, $M_2$, and $M_3$. The most powerful test for the signal sub-hypothesis which has $M_1$ and $M_2$ coming from the same source and $M_3$ being unrelated to them favors the events which have spatial overlap between the localization of $M_1$ and $M_2$, for example the test statistic which is proportional to the product of $M_1$ and $M_2$'s 2D sky (or 3D volume if available) localizations without involving $M_3$'s localization in the integral. However the most powerful test for the corresponding signal sub-hypothesis for $M_1$ and $M_3$ coming from the same source and $M_2$ being unrelated favors the events which have spatial overlap between the localization of $M_1$ and $M_3$. So the two most powerful tests cannot be the same and a UMP test for this search does not exist.

One advantage of dealing with the joint observations of previously individually studied messengers is knowing both their astrophysical and noise originated rate of occurrences. By using these rates, one can empirically weight and combine the sub-hypotheses which has  $l=m$ and also the sub-hypotheses involving multiple same type of messengers coming from the same source. This reduces the number of sub-hypotheses from $f(n+1)$ to $f(n)$. Combining more sub-hypotheses requires assuming rates for multi-messenger observations. Due to the small number of such detections, these rates could not be empirically determined and cannot be used in an objective manner.
\section{Bayesian strategies for models}
\label{sec:bayesian}
As discussed in the previous section, one has to make a model-dependent choice to combine the remaining $f(n)$ sub-hypotheses after using the individual messenger detection rates, which provides the most powerful search for the chosen model. The ratio of the \emph{predicted} number density of multi-messenger sources to the number density of individual messengers' sources together with sources' emission models (i.e. emission energies, dependency on inclination etc.) and messengers' propagations in space, the ratios between the rates of each kind of detection can be found, which is necessary for weighting all of $f(n+1)$ sub-hypotheses. After weighting, null and signal hypotheses reduce to simple hypotheses and the Neyman-Pearson lemma \citep{10.1098/rsta.1933.0009} can be used for finding the most powerful test. The resulting test statistic (TS) is given in Eq. \eqref{ts}.
\begin{equation}
    {\rm TS}(\mathbf{x})=\frac{P(\mathbf{x}|H_s)}{P(\mathbf{x}|H_n)}=\frac{\sum_i P(\mathbf{x}|H_{s}^{i})P(H_{s}^{i})}{\sum_j P(\mathbf{x}|H_{n}^{j})P(H_{n}^{j})} \times \frac{\sum_j P(H_n^j)}{\sum_i P(H_s^i)}
    \label{ts}
\end{equation}
where \(\mathbf{x}\) is the complete set of detection outcomes, $H_s$ and $H_n$ are the signal and null hypotheses in order, and $H_s^i$ and $H_n^j$ are the individual signal and null sub-hypotheses in order. We will ignore the very last term in the Eq. \eqref{ts} as it does not depend on \(\mathbf{x}\). The hypothesis prior probabilities $P(H)$ will be canceled with a same term in detection likelihoods $P(\mathbf{x}|H_a^b)$.

\subsection{Detection likelihoods ${P(\mathbf{x}|H_a^b)}$}
Next, we explain the detection likelihoods. There are two parts to the issue. The first one is the decision of the origin, whether a messenger is astrophysical or noise originated. This part is generally decoupled for different types of messengers due to independent detectors. The detection outcomes for each messenger are used together with the detector characteristics to determine this part. The second one is the multi-messenger aspect of the detection for which correlations between messengers are required, especially in the space-time coordinates of the messengers. This coupling can be done with a source model with parameters ${\boldsymbol \theta}$ as
\begin{equation}
    P(\mathbf{x}|H_a^b)=\int P(\mathbf{x}|\boldsymbol{\theta},H_a^b)P(\boldsymbol{\theta}|H_a^b)d\boldsymbol{\theta}
    \label{eq:integral}
\end{equation}

The source parameters ${\boldsymbol \theta}$ can include any property of the sources (there can be more than one source depending on the sub-hypothesis) such as emission energies or spatial position of the sources. Prior information of such properties can be summarized in a joint density distribution $P(\boldsymbol{\theta})$. If the corresponding sub-hypothesis $H_a^b$ does not include a multi-messenger detection, then there may not be a requirement for a common source and the source parameters ${\boldsymbol \theta}$. In that case, if the detectors are independent from each other, the detection outcomes' probabilities can be expanded as a product.
\begin{equation}
     P(\mathbf{x}|H_a^b)= \prod_i P(\mathbf{x}_i|H_a^b)
     \label{eq:likelihood1}
\end{equation}
where subscript $i$ runs over different detectors and $\mathbf{x}_i$ are the detection outcomes from the $i^{\rm th}$ detector. Similarly, when there is a common source we can expand the detection outcomes' probabilities for a fixed source as a product for different detectors.
\begin{equation}
     P(\mathbf{x}|\{\boldsymbol{\theta}\},H_a^b)= \prod_i P(\mathbf{x}_i|\{\boldsymbol{\theta}\},H_a^b)
     \label{eq:likelihood2}
\end{equation}
There we used the notation $\{\boldsymbol{\theta}\}=\{\boldsymbol{\theta}_1,\boldsymbol{\theta}_2,...\}$ for representing possible set of sources. There can be an additional level of complication related to combinatorics if the sub-hypothesis $H_a^b$ can be satisfied with different groupings of the detections. To illustrate this, consider the sub-hypothesis of having five detected particles, three of them from a source and all the rest being noise originated. In this case the sub-hypothesis can be satisfied with \(\binom{5}{3}=10\) combinations. In such cases we expand the probabilities $P(\mathbf{x}_i|\boldsymbol{\theta},H_a^b)$ as

\begin{multline}
    P(\mathbf{x}_i|\{\boldsymbol{\theta}\},H_a^b)\\=\sum_{\{\mathbf{x}_i^j,\mathbf{x}_i^k,...\}}  P(\mathbf{x}_i|\{\boldsymbol{\theta}\},H_a^b, \{\{\mathbf{x}_i^j,\mathbf{x}_i^k,...\},\{\mathbf{x}_i^p,\mathbf{x}_i^q,...\},...\})\\ \times P( \{\{\mathbf{x}_i^j,\mathbf{x}_i^k,...\},\{\mathbf{x}_i^p,\mathbf{x}_i^q,...\},...\}|\{\boldsymbol{\theta}\},H_a^b)
    \label{eq:likelihood3}
\end{multline}
where the sum is over all the combinations of detection outcomes satisfying the sub-hypothesis $H_a^b$,  the sets $\{\mathbf{x}_i^j,\mathbf{x}_i^k,...\}$ and $\{\mathbf{x}_i^p,\mathbf{x}_i^q,...\}$ are the detection outcomes of individual detections from different sources and $P( \{\{\mathbf{x}_i^j,\mathbf{x}_i^k,...\},\{\mathbf{x}_i^p,\mathbf{x}_i^q,...\},...\}|\{\boldsymbol{\theta}\},H_a^b)$ is equal to the reciprocal of the total possible combinations for $H_a^b$ arising from detector $i$. For example, for a sub-hypothesis with a single source and $w$ particles emitted from that source, and total of $W$ detections; $P(\{\mathbf{x}_i^1,\mathbf{x}_i^2...\mathbf{x}_i^w\}|\boldsymbol{\theta}, H_s^w)=\binom{W}{w}^{-1}$.  $P(\mathbf{x}_i|\{\boldsymbol{\theta}\},H_a^b, \{\{\mathbf{x}_i^j,\mathbf{x}_i^k,...\},\{\mathbf{x}_i^p,\mathbf{x}_i^q,...\},...\})$ are found by physics and empirical data. For memoryless detectors, the detections from different sources are independent so
\begin{multline}
    P(\mathbf{x}_i|\{\boldsymbol{\theta}\},H_a^b, \{\{\mathbf{x}_i^j,\mathbf{x}_i^k,...\},\{\mathbf{x}_i^p,\mathbf{x}_i^q,...\},...\})\\=
    P(\{\mathbf{x}_i^j,\mathbf{x}_i^k,...\}|\boldsymbol{\theta}_1,H_a^b )P(\{\mathbf{x}_i^p,\mathbf{x}_i^q,...\}|\boldsymbol{\theta}_2,H_a^b )
    \label{eq:likelihood4}
\end{multline}
The likelihoods $P(\{\mathbf{x}_i^j,\mathbf{x}_i^k,...\}|\boldsymbol{\theta}_1,H_a^b )$ are found via the detector characteristics and emission models.

Now we look at our second term in Eq. \eqref{eq:integral}, $P(\boldsymbol{\theta}|H_a^b)$. We transform it by using the Bayes' rule.
\begin{equation}
    P(\boldsymbol{\theta}|H_a^b)=\frac{P(H_a^b|\boldsymbol{\theta})P(\boldsymbol{\theta})}{P(H_a^b)}
    \label{eq:bayes}
\end{equation}
The denominator of Eq. \eqref{eq:bayes} cancel with the same term in Eq. \eqref{ts}. $P(\boldsymbol{\theta})$ is the joint density of source parameters being integrated over. 

The sub-hypothesis probabilities $P(H_a^b|\boldsymbol{\theta})$ are found via the expected counts from the sources or the noise origin by assuming a source density and using the empirically known noise trigger rate.

\section{Use cases -- Example: a joint gravitational wave -- high-energy neutrino -- short gamma-ray burst event search}
\label{sec:Use_Cases}
The method for multi-messenger searches introduced above can be used in all scenarios. Specifically, in high-energy astrophysics, one can search for sources which emit more than one messenger. Those messengers can be in any wavelength in the electromagnetic spectrum, and can be neutrinos, cosmic rays, gravitational waves, or any other messenger. Joint emissions of gravitational waves, high-energy neutrinos and short gamma rays from a binary neutron star or a neutron star black hole merger \citep{Kimura_2017,Berger_2014}, or a binary black hole merger in a dense medium, such as an AGN disk, or surrounded by an accretion disk can be such examples \citep{ford2019multimessenger,ford2019agn}. In this section, we examine this example.

Now we give a demonstration of the explained method for three kinds of messengers; gravitational waves (GWs), high-energy neutrinos (neutrinos hereafter), and \textbf{short} gamma-ray bursts (GRBs). We assume a model with continuous single emissions for each messenger type, i.e. no repeated or periodic emission for multi-messenger or single messenger emissions. As mentioned before, there are searches for multi-messenger detection of all the three combinations of two of these messengers \citep{Hamburg_2020,Aartsen_2020,Aartsen_2017}; but there is no triple messenger search. In this search, the start and end times of the GW emission or the gamma-ray emission can be estimated well due to having a continuous detection amplitude, although for neutrinos it is hard to estimate when the emission starts or ends since up to now no continuous cosmic high-energy neutrino flux has been detected. High-energy neutrino emissions are detected in low numbers, generally as a single neutrino. For GWs and GRBs, the detection decision is essentially based on the detected continuous total energy, whereas for neutrinos, it is based on each neutrino's characteristics. Therefore it is more appropriate to separate our signal sub-hypotheses based on different detected neutrino counts (including all the characteristics), i.e., a coincident GW--GRB--$n$ neutrino detection. We will denote our sub-hypotheses with the notation $H_{s_1=\{GW,GRB,a\}, s_2=\{b\}},...,$ where the sets $s_{i}$ in the subscript represent detections from different astrophysical sources. If the set has $GW$ or $GRB$ in it, that means GW or GRB emission was detected from that source. Finally, the sets include positive integers (for example, $a$, $b$), which represent the number of detected high-energy neutrinos from each source. 

For concreteness of the example we consider the ground based interferometric detectors such as LIGO \citep{Aasi2015}, Virgo \citep{Acernese_2014} or KAGRA \citep{kagra2019} for GWs, IceCube \citep{Aartsen_2017_2} for neutrinos and $Fermi$ \citep{Atwood_2009,Meegan_2009} for GRBs. The detection outcomes for GWs are $\mathbf{x}_{GW}=\{t_{GW}, \mathcal{D}, \mathcal{F} \}$ which are the detection time of the GW, its joint volume localization-isotropic equivalent emission energy estimation as a four-dimensional probability distribution, and the estimated false alarm rate. If the joint volume localization-isotropic equivalent emission energy estimation is not explicitly provided; it can be derived from the three-dimensional volume localization, the detected signal energy, and the antenna pattern at the time of the detection. We do not put a constraint on the type of the GW mergers, i.e. binary black hole or neutron star mergers. However, such a distinction can be made by using the mass estimates from the detections too, with a prior. The detection outcomes for high-energy neutrinos are $\mathbf{x}_{\nu}=\{t_{\nu}, \mathbf{\Omega}_{\nu},\sigma_{\nu}, \epsilon_{\nu} \}$, which are the detection times of the neutrinos, their expected sky positions, the angular errors on the sky localizations, and their reconstructed energies. The localization of neutrinos is approximated as two-dimensional Gaussian distribution, and the angular error corresponds to one standard deviation \citep{BRAUN2008299}. The detection outcomes for GRBs are $\mathbf{x}_{\gamma}=\{t_{\gamma}, \mathcal{S} , \mathcal{E}\}$, which are the detection time of the GRB, its localization on the sky and the estimated detected energy. Since we are only considering short gamma ray bursts, we are also implicitly using the duration of the signal such that the analyzed sample's emission durations are $<2$s. Our source parameters are $\boldsymbol \theta=\{{r}_s, \mathbf{\Omega}_{s}, t_s, E_{GW},E_{\nu},E_{\gamma},\kappa_\gamma\}$, which are the distance of the source, its sky position, the retarded reference time of the event (due to the travel time of the messenger), the isotropic equivalent emission energies in GWs, high-energy neutrinos, and gamma rays, and a parameter for relating the peak flux to the total fluence of GRBs. The complete model includes the emission delays of the messengers and the source rates as well, which are explained throughout when they are used. In our analysis we do not use the signalness probabilities provided with the detections, i.e. $p_{astro}$ for GWs or $p_{signalness}$ for neutrinos, since such quantities are Bayesian probabilities and have their own priors; hence are not appropriate to be used in a different Bayesian analysis.
We will first write down the detection likelihoods in Eq. \eqref{eq:likelihood4} which encompasses the ones in Eqs. \eqref{eq:likelihood1} and \eqref{eq:likelihood2}. For a short notation we will denote the sets of detection outcomes from one source $\{\mathbf{x}_i^j,\mathbf{x}_i^k,...\}$ as $\mathbf{x}_{i}$ for each detector.

\subsection{Detection likelihoods}
In this section, for each messenger, we write the detection likelihoods for signal hypotheses with fixed source parameters and for null hypotheses. These likelihoods are used for deducing whether a messenger has astrophysical origin or not, and if it is astrophysical, how likely it is to be associated with the source that has the fixed parameters.

We start with the GWs. The signal likelihood can be expanded as
\begin{multline}
    P(\mathbf{x}_{GW}|\boldsymbol{\theta},H_s)=P(t_{GW},\mathcal{D},\mathcal{F}|t_s,r_s,\mathbf{\Omega}_s,E_{GW},H_s)
    \\=  P(t_{GW}|t_s,H_s)P(\mathcal{F}|t_{GW},r_s,\mathbf{\Omega}_s,E_{GW},H_s)\\ \times P(\mathcal{D}|t_{GW},\mathcal{F},r_s,\mathbf{\Omega}_s,E_{GW},H_s)
\end{multline}

The temporal distribution of $t_{GW}$ is assumed to be uniform around $t_{s}$: $P(t_{GW}|t_s,H_s)=(t_{GW}^{+}-t_{GW}^{-})^{-1}$ for $t_{GW}-t_{s}\in [t_{GW}^{-},t_{GW}^{+}]$ and 0 otherwise. We take $-t_{GW}^{-}=t_{GW}^{+}=250$ s as in \cite{PhysRevD.100.083017,BARET20111}.

By using the Bayes' rule, we expand the likelihood for volume localization.
\begin{multline}
    P(\mathcal{D}|t_{GW},\mathcal{F},r_s,\mathbf{\Omega}_s,E_{GW},H_s)\\=\frac{ P(r_s,\mathbf{\Omega}_s,E_{GW}|t_{GW},\mathcal{D},\mathcal{F},H_s)P(\mathcal{D}|t_{GW},\mathcal{F},H_s)}{P(r_s,\mathbf{\Omega}_s,E_{GW}|t_{GW},\mathcal{F},H_s)}
\end{multline}
The first term in the numerator is the $\mathcal{D}$ distribution itself. We use Bayes' rule for the denominator to have the form 
\begin{multline}
P(\mathcal{D}|t_{GW},r_s,\mathbf{\Omega}_s,E_{GW},\mathcal{F},H_s)\\=\frac{P(\mathcal{F}|t_{GW},H_s)}{P(r_s,\mathbf{\Omega}_s,E_{GW}|t_{GW},H_s)P(\mathcal{F}|t_{GW},r_s,\mathbf{\Omega}_s,E_{GW},H_s)}\\ \times \mathcal{D}(r_s,\mathbf{\Omega}_s,E_{GW})P(\mathcal{D}|t_{GW},\mathcal{F},H_s)
\end{multline}
The term $ P(\mathcal{F}|t_{GW},H_s)$ can be computed by integrating the likelihood for fixed source parameters (which can be obtained from calculations or simulations) over the source parameters. 
\begin{multline}
\label{eq:marginalize}
    P(\mathcal{F}|t_{GW},H_s)=\int  P(\mathcal{F}|t_{GW},r_s,\mathbf{\Omega}_s,E_{GW},H_s)\\ \times P(r_s,\mathbf{\Omega}_s,E_{GW}|t_{GW},H_s)dr_sd\mathbf{\Omega}_sdE_{GW}
\end{multline}
Similarly, for the null hypotheses we expand the likelihood.
\begin{multline}
      P(\mathbf{x}_{GW}|H_n)= P(t_{GW}|H_n)P(\mathcal{F}|t_{GW},H_n)\\ \times P(\mathcal{D}|t_{GW},\mathcal{F},H_n)
\end{multline}
$P(\mathcal{F}|t_{GW},H_n)$ can be found empirically, i.e. through the unphysical time shifted coincidences. We assume the terms  $P(\mathcal{D}|t_{GW},\mathcal{F},H_s)$ and $P(\mathcal{D}|t_{GW},\mathcal{F},H_n)$ do not depend on the hypotheses and are equal to each other, hence cancel in the overall expression. Finally the noise triggers are assumed to be Poisson events and hence can uniformly occur in the observation period $T_{obs}$, $P(t_{GW}|H_n)=T_{obs}^{-1}$. We note that at the end of the full calculation, the end result does not depend on $T_{obs}$; but we do not drop it throughout for clarity.

Next we move on the signal likelihoods for neutrinos and expand similarly.
\begin{multline}
\label{eq:HENsignallikelihood}
    P(t_{\nu},\epsilon_{\nu},\sigma_{\nu},\mathbf{\Omega}_{\nu}|\boldsymbol{\theta}, H_s)=P(t_{\nu}|t_s,H_s) P(\epsilon_{\nu}|\mathbf{\Omega}_{s},H_s)\\ \times P(\mathbf{\Omega}_{\nu},\sigma_{\nu}|\epsilon_{\nu},\mathbf{\Omega}_{s},H_s)
\end{multline}

The temporal distribution of $t_{\nu}$ is also assumed to be uniform around $t_{s}$: $P(t_{\nu}|t_s,H_s)=(t_{\nu}^{+}-t_{\nu}^{-})^{-1}$ for $t_{\nu}-t_{s}\in [t_{\nu}^{-},t_{\nu}^{+}]$ and 0 otherwise. We take $-t_{\nu}^{-}=t_{\nu}^{+}=250$ s as in \cite{PhysRevD.100.083017,BARET20111}.

The estimated source localization from the detection can be written as
\begin{equation}
    P(\mathbf{\Omega}_{s}|\epsilon_{\nu},\sigma_{\nu},\mathbf{\Omega}_{\nu}, H_s)=\frac{e^{\frac{-|\mathbf{\Omega}_{\nu}-\mathbf{\Omega}_{s}|^2}{2\sigma_{\nu}^2}}}{2\pi \sigma_{\nu}^2}
\end{equation}
However we need the probability $P(\sigma_{\nu},\mathbf{\Omega}_{\nu}|\epsilon_{\nu},\mathbf{\Omega}_{s}, H_s)$ which we expand with Bayes' rule as
\begin{multline}
\label{eq:HENsignallikelihood2}
    P(\sigma_{\nu},\mathbf{\Omega}_{\nu}|\epsilon_{\nu},\mathbf{\Omega}_{s}, H_s)\\=\frac{P(\mathbf{\Omega}_{s}|\epsilon_{\nu},\sigma_{\nu},\mathbf{\Omega}_{\nu}, H_s)P(\sigma_{\nu},\mathbf{\Omega}_{\nu}|\epsilon_{\nu},H_s)}{P(\mathbf{\Omega}_{s}|\epsilon_{\nu},H_s)}\\=\frac{e^{\frac{-|\mathbf{\Omega}_{\nu}-\mathbf{\Omega}_{s}|^2}{2\sigma_{\nu}^2}}}{2\pi \sigma_{\nu}^2}\frac{P(\sigma_{\nu},\mathbf{\Omega}_{\nu}|\epsilon_{\nu},H_s)}{P(\mathbf{\Omega}_{s}|\epsilon_{\nu},H_s)}\\=P(\sigma_{\nu},\mathbf{\Omega}_{\nu}|\epsilon_{\nu},H_s)\frac{e^{\frac{-|\mathbf{\Omega}_{\nu}-\mathbf{\Omega}_{s}|^2}{2\sigma_{\nu}^2}}P(\epsilon_{\nu}|H_s)}{2\pi \sigma_{\nu}^2 P(\epsilon_{\nu}|\mathbf{\Omega}_{s},H_s)P(\mathbf{\Omega}_{s}|H_s)}
\end{multline}
By assuming a power law with exponent -2 for the energy distribution of neutrinos \citep{1997PhRvL..78.2292W} and by using the effective area of the neutrino detector  $A_{eff}(\epsilon_{\nu},\mathbf{\Omega}_{s})$ we write the term $P(\epsilon_{\nu}|H_s)$ as
\begin{equation}
    P(\epsilon_{\nu}|H_s)=\frac{\int A_{eff}(\epsilon_{\nu},\mathbf{\Omega}_{s})\epsilon_{\nu}^{-2}P(\mathbf{\Omega}_{s}|H_s) d\mathbf{\Omega}_{s}}{\int_{\epsilon_{min}}^{\epsilon_{max}} \int A_{eff}(\epsilon_{\nu}',\mathbf{\Omega}_{s})\epsilon_{\nu}'^{-2}P(\mathbf{\Omega}_{s}|H_s) d\mathbf{\Omega}_{s}d\epsilon_{\nu}'}
\end{equation}
$\epsilon_{min},\epsilon_{max}$ are 100 GeV and 100 PeV for IceCube.
The $P(\epsilon_{\nu}|\mathbf{\Omega}_{s},H_s)$ terms in Eqs. \eqref{eq:HENsignallikelihood} and \eqref{eq:HENsignallikelihood2} cancel. 

Next, we expand the null hypothesis likelihood similarly.
\begin{multline}
    P(t_{\nu},\epsilon_{\nu},\sigma_{\nu},\mathbf{\Omega}_{\nu}| H_n)=P(t_{\nu}|H_n)P(\epsilon_{\nu},\sigma_{\nu},\mathbf{\Omega}_{\nu}| t_{\nu},H_n) \\= P(t_{\nu}| H_n)P(\epsilon_{\nu}| t_{\nu},H_n)P(\sigma_{\nu},\mathbf{\Omega}_{\nu}|\epsilon_{\nu}, t_{\nu},H_n)
\end{multline}
$P(t_{\nu}|H_n)=T_{obs}^{-1}$ and $P(\epsilon_{\nu}| t_{\nu},H_n)$ can be found empirically from detector characteristics and past observations. The time dependency of the last term comes from the annual modulation due to Earth's motion around the Sun and can be expressed with a function $\mathcal{T}(t_{\nu},\epsilon_{\nu},\mathbf{\Omega}_{\nu})$ whose average over one year for every $(\epsilon_{\nu},\mathbf{\Omega}_{\nu})$ pair is one.
\begin{equation}
    P(\sigma_{\nu},\mathbf{\Omega}_{\nu}|\epsilon_{\nu}, t_{\nu},H_n)=P(\sigma_{\nu},\mathbf{\Omega}_{\nu}|\epsilon_{\nu},H_n)\mathcal{T}(t_{\nu},\epsilon_{\nu},\mathbf{\Omega}_{\nu})
\end{equation}
The terms $P(\sigma_{\nu},\mathbf{\Omega}_{\nu}|\epsilon_{\nu},H_s)$ and $P(\sigma_{\nu},\mathbf{\Omega}_{\nu}|\epsilon_{\nu},H_n)$ do not depend on the hypotheses and cancel in the overall expression. 

Third, we move on the likelihoods for GRBs and expand similarly.
\begin{multline}
\label{eq:GRBsignallikelihood}
    P(t_{\gamma},\mathcal{S},\mathcal{E}|\boldsymbol{\theta},H_s)=P(t_{\gamma}|t_{s},H_s)\\ \times P(\mathcal{S},\mathcal{E}|t_{\gamma},\mathbf{\Omega}_{s},r_s,E_{\gamma},\kappa_\gamma,H_s)
\end{multline}
The temporal distribution of $t_{\gamma}$ is also assumed to be uniform around $t_{s}$: $P(t_{\gamma}|t_s,H_s)=(t_{\gamma}^{+}-t_{\gamma}^{-})^{-1}$ for $t_{\gamma}-t_{s}\in [t_{\gamma}^{-},t_{\gamma}^{+}]$ and 0 otherwise. We take $t_{\gamma}^{-}=100$ s and $t_{\gamma}^{+}=250$ s from \cite{BARET20111}.
For the second term in the likelihood we again use the Bayes' rule.
\begin{multline}
    P(\mathcal{S},\mathcal{E}|t_{\gamma},\mathbf{\Omega}_{s},r_s,E_{\gamma},\kappa_\gamma,H_s) \\=P(\mathbf{\Omega}_{s},r_s,E_{\gamma}|\mathcal{S},\mathcal{E},t_{\gamma},\kappa_\gamma,H_s) \\ \times \frac{P(\mathcal{E}|t_{\gamma},\kappa_\gamma,H_s)P(\mathcal{S}|\mathcal{E},t_{\gamma},\kappa_\gamma,H_s)}{P(\mathbf{\Omega}_{s},r_s,E_{\gamma}|t_{\gamma},\kappa_\gamma,H_s)}
\end{multline}
The first term is the position and energy estimations themselves
\begin{equation}
    P(\mathbf{\Omega}_{s},r_s,E_{\gamma}|\mathcal{S},\mathcal{E},t_{\gamma},\kappa_\gamma,H_s)=\mathcal{S}(\mathbf{\Omega}_{s})\delta(\mathcal{E}-\eta \frac{E_{\gamma}}{4\pi r_s^2})
\end{equation}
where $\eta$ is a constant describing the detection efficiency of the detector. $P(\mathcal{E}|t_{\gamma},\kappa_\gamma,H_s)$ term can be computed by marginalizing the conditional probability with fixed source parameters over the source parameters just like in Eq. \eqref{eq:marginalize} for the GWs.
\begin{multline}
    P(\mathcal{E}|t_{\gamma},\kappa_\gamma,H_s)=\int P(\mathcal{E}|t_{\gamma},\mathbf{\Omega}_{s},r_s,E_{\gamma},\kappa_\gamma,H_s)\\ \times P(\mathbf{\Omega}_{s},r_s,E_{\gamma}|t_{\gamma},\kappa_\gamma,H_s)d\mathbf{\Omega}_{s}dr_sdE_{\gamma}
\end{multline}

For $P(\mathcal{S}|\mathcal{E},t_{\gamma},\kappa_\gamma,H_s)$ term we ignore the effect of the peak flux to total fluence ratio ($\kappa_\gamma$), which is the case especially for refined Human in the Loop (HitL) localizations \citep{Connaughton_2015}.
\begin{equation}
    P(\mathcal{S}|\mathcal{E},t_{\gamma},\kappa_\gamma,H_s)=P(\mathcal{S}|\mathcal{E},t_{\gamma},H_s)
\end{equation}

We expand the null hypothesis likelihoods as
\begin{multline}
    P(t_{\gamma},\mathcal{S},\mathcal{E}|H_n)=P(t_{\gamma}|H_n)\\
    \times P(\mathcal{E}|t_{\gamma},H_n)P(\mathcal{S}|t_{\gamma},\mathcal{E},H_n)
\end{multline}
$P(\mathcal{S}|t_{\gamma},\mathcal{E},H_s)$ and $P(\mathcal{S}|t_{\gamma},\mathcal{E},H_n)$ terms do not depend on hypotheses and cancel in the overall expression. $P(\mathcal{E}|t_{\gamma},H_n)$ can be found via the noise characteristics of the detector and $P(t_{\gamma}|H_n)=T_{obs}^{-1}$.

\subsection{Prior sub-hypothesis probabilities}
Now we move on the prior probabilities for each sub-hypothesis. These are found by assuming each detection candidate trigger (noise or astrophysical origin) is a Poisson event. The expected counts for the Poisson processes are found by the known noise trigger rates $R_{bg, \xi}$ and the \emph{assumed true} astrophysical source rates $\Dot{n}_{\xi}^{true}$ for the messenger $\xi$. We are interested in the \emph{observable} source rates $\Dot{n}_{\xi}$ for GWs and GRBs which have detection cuts in terms of the signal to noise power ratio or photon count. We define $\rho(\frac{E_{GW}}{r_{s}^2},\mathbf{\Omega}_{s},t_{s})$ and $I(\frac{E_{\gamma}}{r_{s}^2},\mathbf{\Omega}_{s},t_{s},\kappa_\gamma)$ functions as the cut functions and the detection thresholds $\rho_{th}$ and $I_{th}$. $\rho$ can be taken as the network signal-to-noise ratio for GWs and $I$ as the detected peak flux. Those functions take into account the effective antenna pattern of the GW detector network (by accounting the different sensitivities of the detectors too) and the view of the $Fermi$ satellite. In order to calculate the peak flux from the fluence in function $I$, one needs to assume an emission form as well. For this purpose, the distribution of peak flux to total fluence ratio ($\kappa_\gamma$) can be taken from previous measurements and can be additionally marginalized over. Furthermore, we assume beaming for neutrino and gamma-ray emission from the same opening with a beaming factor $f_b\ (\sim 10-100)$. The observable source rate for a source emitting only GWs is
\begin{equation}
    \Dot{n}_{GW}=\int \Dot{n}_{GW}^{true}P(\boldsymbol{\theta})[\rho(\frac{E_{GW}}{r_{s}^2},\mathbf{\Omega}_{s},t_{s})\geq\rho_{th} ] d\boldsymbol{\theta}
\end{equation}
The binary bracket notation $[\zeta]$ is 1 if $\zeta$ is true and 0 if false. For a GRB only source
\begin{equation}
    \Dot{n}_{\gamma}=f_b^{-1}\int \Dot{n}_{\gamma}^{true}P(\boldsymbol{\theta})[I(\frac{E_{\gamma}}{r_{s}^2},\mathbf{\Omega}_{s},t_{s},\kappa_\gamma)\geq I_{th} ] d\boldsymbol{\theta}
\end{equation}
For a multi-messenger source it is
\begin{multline}
    \Dot{n}_{GW,\gamma}=f_b^{-1}\int \Dot{n}_{GW,\gamma}^{true}P(\boldsymbol{\theta})[\rho(\frac{E_{GW}}{r_{s}^2},\mathbf{\Omega}_{s},t_{s})\geq\rho_{th} ]\\ \times [I(\frac{E_{\gamma}}{r_{s}^2},\mathbf{\Omega}_{s},t_{s},\kappa_\gamma)\geq I_{th} ] d\boldsymbol{\theta}
\end{multline}
For neutrinos we are interested in the observable neutrino rate rather than the observable source rate.
\begin{equation}
    \Dot{n}_{\nu}=f_b^{-1}\int \Dot{n}_{\nu}^{true}P(\boldsymbol{\theta})\langle n_{\nu}(E_{\nu},r_{s},\mathbf{\Omega}_{s})\rangle d\boldsymbol{\theta}
\end{equation}
$\langle n_{\nu}(E_{\nu},r_{s},\mathbf{\Omega}_{s})\rangle$ is the detector specific expected number of detected neutrinos from a source with given location and emission energy which scales linearly with $\frac{E_{\nu}}{r_{s}^2}$ and depends on the effective area. For a multi-messenger detection with neutrinos the interesting quantity would be
\begin{multline}
    \Dot{n}_{GW,\nu,\gamma}=f_b^{-1}\int \Dot{n}_{GW,\nu,\gamma}^{true}P(\boldsymbol{\theta})\langle n_{\nu}(E_{\nu},r_{s},\mathbf{\Omega}_{s})\rangle\\ \times [\rho(\frac{E_{GW}}{r_{s}^2},\mathbf{\Omega}_{s},t_{s})\geq\rho_{th} ] [I(\frac{E_{\gamma}}{r_{s}^2},\mathbf{\Omega}_{s},t_{s},\kappa_\gamma)\geq I_{th} ] d\boldsymbol{\theta}
\end{multline}

For clarity, let's demonstrate a specific sub-hypothesis $H_{s_1=\{GW,GRB,a\},s_2=\{b\}}$ for detected $\alpha$ GWs, $\beta$ GRBs and $\mu$ neutrinos in total. As a reminder, that sub-hypothesis corresponds to signal detections from two sources; from the first one, $s_1$, a GW, a GRB and $a$ neutrinos are detected, and from the second one, $s_2$, only $b$ neutrinos are detected. We will denote the probability of occurrence of $d$ Poisson events with an expectation $\lambda$ as $Poi(d,\lambda)=\frac{\lambda^d e^{-\lambda}}{d!}$. In this sub-hypothesis the first source is clearly a multi-messenger source; but the second one can be a multi-messenger source from which the GW or the GRB or both were not detected, or it can be simply a source which only emits neutrinos. We consider all of these four possible cases.
\begin{widetext}
\begin{multline}
\label{eq:prior}
    P(H_{s_1=\{GW,GRB,a\},s_2=\{b\}}|\boldsymbol{\theta}_1,\boldsymbol{\theta}_2)=  Poi(\mu-a-b,R_{bg,\nu}T_{obs}) Poi(\alpha-1,R_{bg,GW}T_{obs})Poi(\beta-1,R_{bg,\gamma}T_{obs})\\ \times
    Poi(a,\langle n_{\nu}(E_{\nu_1},r_{s_1},\mathbf{\Omega}_{s_1})\rangle)
    [\rho(\frac{E_{GW_1}}{r_{s_1}^2},\mathbf{\Omega}_{s_1},t_{s_1})\geq\rho_{th} ]
    [I(\frac{E_{\gamma_1}}{r_{s_1}^2},\mathbf{\Omega}_{s_1},t_{s_1},\kappa_\gamma)\geq I_{th} ] Poi(b,\langle n_{\nu}(E_{\nu_2},r_{s_2},\mathbf{\Omega}_{s_2})\rangle)
    \\ \times 
    \{Poi(2,\Dot{n}_{GW,\nu,\gamma}T_{obs})Poi(0,(\Dot{n}_{GW}-\Dot{n}_{GW,\nu,\gamma})T_{obs}) Poi(0,(\Dot{n}_{\nu}-\Dot{n}_{GW,\nu,\gamma})T_{obs})Poi(0,(\Dot{n}_{\gamma}-\Dot{n}_{GW,\nu,\gamma})T_{obs})\\ \times [\rho(\frac{E_{GW_2}}{r_{s_2}^2},\mathbf{\Omega}_{s_2},t_{s_2})<\rho_{th} ]
    [I(\frac{E_{\gamma_2}}{r_{s_2}^2},\mathbf{\Omega}_{s_2},t_{s_2},\kappa_\gamma)<I_{th} ] 
    \\+Poi(1,\Dot{n}_{GW,\nu,\gamma}T_{obs})Poi(1,\Dot{n}_{GW,\nu}T_{obs}) Poi(0,(\Dot{n}_{GW}-\Dot{n}_{GW,\nu}-\Dot{n}_{GW,\nu,\gamma})T_{obs})\\ \times Poi(0,(\Dot{n}_{\nu}-\Dot{n}_{GW,\nu}-\Dot{n}_{GW,\nu,\gamma})T_{obs}) Poi(0,(\Dot{n}_{\gamma}-\Dot{n}_{GW,\nu,\gamma})T_{obs})
    [\rho(\frac{E_{GW_2}}{r_{s_2}^2},\mathbf{\Omega}_{s_2},t_{s_2})<\rho_{th} ]
    \\ 
    +Poi(1,\Dot{n}_{GW,\nu,\gamma}T_{obs})Poi(1,\Dot{n}_{\nu,\gamma}T_{obs}) Poi(0,(\Dot{n}_{GW}-\Dot{n}_{GW,\nu,\gamma})T_{obs}) Poi(0,(\Dot{n}_{\nu}-\Dot{n}_{\nu,\gamma}-\Dot{n}_{GW,\nu,\gamma})T_{obs})\\ \times Poi(0,(\Dot{n}_{\gamma}-\Dot{n}_{\nu,\gamma}-\Dot{n}_{GW,\nu,\gamma})T_{obs})
    [I(\frac{E_{\gamma_2}}{r_{s_2}^2},\mathbf{\Omega}_{s_2},t_{s_2},\kappa_\gamma)<I_{th} ]
    \\+Poi(1,\Dot{n}_{GW,\nu,\gamma}T_{obs})Poi(0,(\Dot{n}_{GW}-\Dot{n}_{GW,\nu,\gamma})T_{obs}) Poi(1,(\Dot{n}_{\nu}-\Dot{n}_{GW,\nu,\gamma})T_{obs})Poi(0,(\Dot{n}_{\gamma}-\Dot{n}_{GW,\nu,\gamma})T_{obs}) 
     \}
\end{multline}
\end{widetext}
\subsection{Source parameter distributions}
Finally, we explain the required distributions for source parameters. First, we write the complete distribution. The sources are distributed such that the event rate is uniform in the comoving spacetime. There can be many different models for the emission energies. Here we provide only a naive example. We assume log uniform distributions for GW, neutrino, and GRB emission energies \citep{PhysRevD.100.083017}. We take the limits of neutrino and GRB emissions to be $10^{47}-10^{52}$ erg \citep{Veske_2020,Aartsen_2020,Berger_2014,Abbott_2017_2} and GW limits to be between $0.1-10$ $M_{\odot}c^2$ (assuming $\sim 5\%$ of the mass is emitted in a merger \citep{Abbott_2019}). The event time is distributed uniformly in the observation time. GRBs have chaotic forms, therefore the peak flux to total fluence ratio cannot be modeled well. Simply $P(\kappa_\gamma)$ can be taken as the reciprocal distribution of the durations.
\begin{equation}
    P(\boldsymbol{\theta})=\frac{P(\kappa_\gamma) r_s^2}{4\pi (1+z(r_s))^4 E_{GW}E_{\nu}E_{\gamma} T_{obs} N_r log(100)^3 }
\end{equation}
$N_r$ is the normalization constant for $r_s$. $z(r_s)$ is the redshift and the factor $(1+z(r_s))^4$ in the denominator accounts for the dilution of sources in space and the time dilation due to Hubble expansion.

There are three conditional source distributions used in the likelihoods. The one in the GW part is
\begin{equation}
    P(r_s,\mathbf{\Omega}_s,E_{GW}|t_{GW},H_s)=\frac{\frac{r_s^2}{4\pi}[\rho(\frac{E_{GW}}{r_{s}^2},\mathbf{\Omega}_{s},t_{GW})\geq\rho_{th}]}{ (1+z(r_s))^4 E_{GW} N_r' log(100) } 
\end{equation}
$N_r'$ is the normalization constant. Here we ignored the effect of using $t_{GW}$ instead of $t_s$ in the $\rho$ function. For greater accuracy a new function can also be defined.

The conditional distribution in the neutrino part is
\begin{equation}
    P(\mathbf{\Omega}_s|H_s)= \frac{\int_{\epsilon_{min}}^{\epsilon_{max}} A_{eff}(\epsilon_{\nu},\mathbf{\Omega}_{s})\epsilon_{\nu}^{-2}d\epsilon_{\nu} }{ \int \int_{\epsilon_{min}}^{\epsilon_{max}} A_{eff}(\epsilon_{\nu},\mathbf{\Omega}_{s}')\epsilon_{\nu}^{-2}d\epsilon_{\nu}d\mathbf{\Omega}_s'}
\end{equation}

The conditional distribution in the GRB part is
\begin{equation}
    P(r_s,\mathbf{\Omega}_s,E_{\gamma}|t_{\gamma},\kappa_\gamma,H_s)=\frac{r_s^2[I(f_b\frac{E_{\gamma}}{r_{s}},\mathbf{\Omega}_{s},t_{\gamma},\kappa_\gamma)\geq I_{th}]}{4\pi (1+z(r_s))^4 E_{\gamma} N_r'' log(100) } 
\end{equation}
$N_r''$ is the normalization constant. Here we also ignored the effect of using $t_{\gamma}$ instead of $t_s$ in the $I$ function. For greater accuracy a new function can also be defined.

With the guidance provided in this section, a realtime multi-messenger search for GWs, neutrinos and GRBs can be constructed.

\section{Conclusion}
\label{sec:conclusion}
In this paper, we addressed the problem of optimal multi-messenger searches. Having more messengers will not only make us better understand their sources; but can also increase the significance of sub-threshold single messenger detections and increase the rate of detections without a necessary upgrade to the detectors. 

We showed that a model-independent optimal solution does not exist. We provided a Bayesian solution that is scalable to any number of messengers. It is based on constructing a test statistic by combining different sub-hypotheses via using their predicted rates according to a model. This gives the highest power for the regular frequentist hypothesis test for the assumed model. As a Bayesian solution, this method's performance is dependent on the accuracy of the current models. The described method is completely scalable and applicable to any number and any kind of messengers.

Finally, we examined the use case for a search for joint GW-neutrino-GRB emissions. Although there are searches for all the three combinations of two of these messengers \citep{Hamburg_2020,Aartsen_2020,Aartsen_2017}, this is the first examination of the triple messenger search, which can be applied in real time e.g., similarly to \cite{2019arXiv190105486C,Keivani:2019smf}.
\section*{Acknowledgments}
The authors thank Benjamin Farr and Gregory Ashton for useful feedback. The authors are thankful for the support of Columbia University in the City of New York and the National Science Foundation grant PHY-2012035. D.V. acknowledges Jacob Shaham Fellowship. I.B. acknowledges support from the National Science Foundation under grant PHY-1911796 and the Alfred P. Sloan Research Foundation. This document was reviewed by the LIGO Scientific Collaboration under the document number P2000377.

\bibliography{references} 

\begin{thebibliography}{}
\expandafter\ifx\csname natexlab\endcsname\relax\def\natexlab#1{#1}\fi
\providecommand{\url}[1]{\href{#1}{#1}}
\providecommand{\dodoi}[1]{doi:~\href{http://doi.org/#1}{\nolinkurl{#1}}}
\providecommand{\doeprint}[1]{\href{http://ascl.net/#1}{\nolinkurl{http://ascl.net/#1}}}
\providecommand{\doarXiv}[1]{\href{https://arxiv.org/abs/#1}{\nolinkurl{https://arxiv.org/abs/#1}}}

\bibitem[{Aartsen {et~al.}(2017{\natexlab{a}})Aartsen, Ackermann, Adams,
  Aguilar, Ahlers, Ahrens, Altmann, Andeen, Anderson, Ansseau,
  {et~al.}}]{Aartsen_2017_2}
Aartsen, M., Ackermann, M., Adams, J., {et~al.} 2017{\natexlab{a}}, Journal of
  Instrumentation, 12, P03012, \dodoi{10.1088/1748-0221/12/03/p03012}

\bibitem[{Aartsen {et~al.}(2017{\natexlab{b}})Aartsen, Ackermann, Adams,
  Aguilar, Ahlers, Ahrens, Samarai, Altmann, Andeen, Anderson, \&
  et~al.}]{Aartsen_2017}
Aartsen, M.~G., Ackermann, M., Adams, J., {et~al.} 2017{\natexlab{b}}, The
  Astrophysical Journal, 843, 112, \dodoi{10.3847/1538-4357/aa7569}

\bibitem[{Aartsen {et~al.}(2018)Aartsen, Ackermann, Adams, Aguilar, Ahlers,
  Ahrens, Samarai, Altmann, Andeen, Anderson, {et~al.}}]{blazar}
---. 2018, Science, 361, 147

\bibitem[{Aartsen {et~al.}(2020)Aartsen, Ackermann, Adams, Aguilar, Ahlers,
  Ahrens, Alispach, Andeen, Anderson, Ansseau, \& et~al.}]{Aartsen_2020}
---. 2020, The Astrophysical Journal Letters, 898, L10,
  \dodoi{10.3847/2041-8213/ab9d24}

\bibitem[{Aasi {et~al.}(2015)Aasi, Abbott, Abbott, Abbott, Abernathy, Ackley,
  Adams, Adams, Addesso, {et~al.}}]{Aasi2015}
Aasi, J., Abbott, B.~P., Abbott, R., {et~al.} 2015, Classical and Quantum
  Gravity, 32, 074001, \dodoi{10.1088/0264-9381/32/7/074001}

\bibitem[{Abbott {et~al.}(2019)Abbott, Abbott, Abbott, Abraham, Acernese,
  Ackley, Adams, Adhikari, Adya, Affeldt, \& et~al.}]{Abbott_2019}
Abbott, B., Abbott, R., Abbott, T., {et~al.} 2019, Physical Review X, 9,
  031040, \dodoi{10.1103/physrevx.9.031040}

\bibitem[{Abbott {et~al.}(2016)Abbott, Abbott, Abbott, Abernathy, Acernese,
  Ackley, Adams, Adams, Addesso, Adhikari, {et~al.}}]{PhysRevLett.116.061102}
Abbott, B.~P., Abbott, R., Abbott, T.~D., {et~al.} 2016, Phys. Rev. Lett., 116,
  061102, \dodoi{10.1103/PhysRevLett.116.061102}

\bibitem[{Abbott {et~al.}(2017{\natexlab{a}})Abbott, Abbott, Abbott, Acernese,
  Ackley, Adams, Adams, Addesso, Adhikari, Adya, {et~al.}}]{Abbott_2017}
---. 2017{\natexlab{a}}, The Astrophysical Journal Letters, 848, L12,
  \dodoi{10.3847/2041-8213/aa91c9}

\bibitem[{Abbott {et~al.}(2017{\natexlab{b}})Abbott, Abbott, Abbott, Acernese,
  Ackley, Adams, Adams, Addesso, Adhikari, Adya, \& et~al.}]{Abbott_2017_2}
---. 2017{\natexlab{b}}, The Astrophysical Journal Letters, 848, L13,
  \dodoi{10.3847/2041-8213/aa920c}

\bibitem[{Acernese {et~al.}(2014)Acernese, Agathos, Agatsuma, Aisa, Allemandou,
  Allocca, Amarni, Astone, Balestri, {et~al.}}]{Acernese_2014}
Acernese, F., Agathos, M., Agatsuma, K., {et~al.} 2014, Classical and Quantum
  Gravity, 32, 024001, \dodoi{10.1088/0264-9381/32/2/024001}

\bibitem[{Akutsu {et~al.}(2019)Akutsu, Ando, Arai, Arai, Araki, Araya, Aritomi,
  Asada, Aso, Atsuta, {et~al.}}]{kagra2019}
Akutsu, T., Ando, M., Arai, K., {et~al.} 2019, Nature Astronomy, 3, 35–40,
  \dodoi{10.1038/s41550-018-0658-y}

\bibitem[{Arnett {et~al.}(1989)Arnett, Bahcall, Kirshner, \&
  Woosley}]{doi:10.1146/annurev.aa.27.090189.003213}
Arnett, W.~D., Bahcall, J.~N., Kirshner, R.~P., \& Woosley, S.~E. 1989, Annual
  Review of Astronomy and Astrophysics, 27, 629,
  \dodoi{10.1146/annurev.aa.27.090189.003213}

\bibitem[{Ashton {et~al.}(2018)Ashton, Burns, Canton, Dent, Eggenstein,
  Nielsen, Prix, Was, \& Zhu}]{Ashton_2018}
Ashton, G., Burns, E., Canton, T.~D., {et~al.} 2018, The Astrophysical Journal,
  860, 6, \dodoi{10.3847/1538-4357/aabfd2}

\bibitem[{Aso {et~al.}(2008)Aso, M{\'{a}}rka, Finley, Dwyer, Kotake, \&
  M{\'{a}}rka}]{2008CQGra..25k4039A}
Aso, Y., M{\'{a}}rka, Z., Finley, C., {et~al.} 2008, Class. Quantum Grav, 25,
  114039, \dodoi{10.1088/0264-9381/25/11/114039}

\bibitem[{Atwood {et~al.}(2009)Atwood, Abdo, Ackermann, Althouse, Anderson,
  Axelsson, Baldini, Ballet, Band, Barbiellini, \& et~al.}]{Atwood_2009}
Atwood, W.~B., Abdo, A.~A., Ackermann, M., {et~al.} 2009, The Astrophysical
  Journal, 697, 1071–1102, \dodoi{10.1088/0004-637x/697/2/1071}

\bibitem[{Baret {et~al.}(2011)Baret, Bartos, Bouhou, Corsi, Palma, Donzaud,
  Elewyck, Finley, Jones, Kouchner, Márka, Márka, Moscoso, Chassande-Mottin,
  Papa, Pradier, Raffai, Rollins, \& Sutton}]{BARET20111}
Baret, B., Bartos, I., Bouhou, B., {et~al.} 2011, Astroparticle Physics, 35, 1
  , \dodoi{https://doi.org/10.1016/j.astropartphys.2011.04.001}

\bibitem[{Baret {et~al.}(2012)Baret, Bartos, Bouhou, Chassande-Mottin, Corsi,
  Di~Palma, Donzaud, Drago, Finley, Jones, Klimenko, Kouchner, M\'arka,
  M\'arka, Moscoso, Papa, Pradier, Prodi, Raffai, Re, Rollins, Salemi, Sutton,
  Tse, Van~Elewyck, \& Vedovato}]{PhysRevD.85.103004}
---. 2012, Phys. Rev. D, 85, 103004, \dodoi{10.1103/PhysRevD.85.103004}

\bibitem[{Bartos {et~al.}(2019)Bartos, Veske, Keivani, M\'arka, Countryman,
  Blaufuss, Finley, \& M\'arka}]{PhysRevD.100.083017}
Bartos, I., Veske, D., Keivani, A., {et~al.} 2019, Phys. Rev. D, 100, 083017,
  \dodoi{10.1103/PhysRevD.100.083017}

\bibitem[{Berger(2014)}]{Berger_2014}
Berger, E. 2014, Annual Review of Astronomy and Astrophysics, 52, 43–105,
  \dodoi{10.1146/annurev-astro-081913-035926}

\bibitem[{Braun {et~al.}(2008)Braun, Dumm, {De Palma}, Finley, Karle, \&
  Montaruli}]{BRAUN2008299}
Braun, J., Dumm, J., {De Palma}, F., {et~al.} 2008, Astroparticle Physics, 29,
  299 , \dodoi{https://doi.org/10.1016/j.astropartphys.2008.02.007}

\bibitem[{Connaughton {et~al.}(2015)Connaughton, Briggs, Goldstein, Meegan,
  Paciesas, Preece, Wilson-Hodge, Gibby, Greiner, Gruber, Jenke, Kippen,
  Pelassa, Xiong, Yu, Bhat, Burgess, Byrne, Fitzpatrick, Foley, Giles, Guiriec,
  van~der Horst, von Kienlin, McBreen, McGlynn, Tierney, \&
  Zhang}]{Connaughton_2015}
Connaughton, V., Briggs, M.~S., Goldstein, A., {et~al.} 2015, The Astrophysical
  Journal Supplement Series, 216, 32, \dodoi{10.1088/0067-0049/216/2/32}

\bibitem[{{Countryman} {et~al.}(2019){Countryman}, {Keivani}, {Bartos},
  {Marka}, {Kintscher}, {Corley}, {Blaufuss}, {Finley}, \&
  {Marka}}]{2019arXiv190105486C}
{Countryman}, S., {Keivani}, A., {Bartos}, I., {et~al.} 2019, arXiv e-prints,
  arXiv:1901.05486.
\newblock \doarXiv{1901.05486}

\bibitem[{Davis {et~al.}(1968)Davis, Harmer, \& Hoffman}]{PhysRevLett.20.1205}
Davis, R., Harmer, D.~S., \& Hoffman, K.~C. 1968, Phys. Rev. Lett., 20, 1205,
  \dodoi{10.1103/PhysRevLett.20.1205}

\bibitem[{{Figueiredo} {et~al.}(1990){Figueiredo}, {Villela}, {Jayanthi},
  {Wuensche}, {Neri}, \& {Cesta}}]{1990RMxAA..21..459F}
{Figueiredo}, N., {Villela}, T., {Jayanthi}, U.~B., {et~al.} 1990, Revista
  Mexicana de Astronomia y Astrofisica, 21, 459

\bibitem[{Ford {et~al.}(2019)Ford, Fraschetti, Fryer, Liebling, Perna, Shawhan,
  Veres, \& Zhang}]{ford2019multimessenger}
Ford, K. E.~S., Fraschetti, F., Fryer, C., {et~al.} 2019.
\newblock \doarXiv{1903.11116}

\bibitem[{{Ford} {et~al.}(2019){Ford}, {Bartos}, {McKernan}, {Haiman}, {Corsi},
  {Keivani}, {Marka}, {Perna}, {Graham}, {Ross}, {Stern}, {Bellovary}, {Berti},
  {O'Dowd}, {Lyra}, {MacLow}, \& {Marka}}]{ford2019agn}
{Ford}, K.~E.~S., {Bartos}, I., {McKernan}, B., {et~al.} 2019, \baas, 51, 247.
\newblock \doarXiv{1903.09529}

\bibitem[{Giacconi(2003)}]{RevModPhys.75.995}
Giacconi, R. 2003, Rev. Mod. Phys., 75, 995, \dodoi{10.1103/RevModPhys.75.995}

\bibitem[{Hamburg {et~al.}(2020)Hamburg, Fletcher, Burns, Goldstein, Bissaldi,
  Briggs, Cleveland, Giles, Hui, Kocevski, {et~al.}}]{Hamburg_2020}
Hamburg, R., Fletcher, C., Burns, E., {et~al.} 2020, The Astrophysical Journal,
  893, 100, \dodoi{10.3847/1538-4357/ab7d3e}

\bibitem[{{HAWC Collaboration}(2019)}]{hawc}
{HAWC Collaboration}. 2019, LIGO/Virgo S191216ap: HAWC gamma-ray sub-threshold
  event coincident with LIGO/Virgo and IceCube localizations, GCN Circular
  26472.
\newblock \url{gcn.gsfc.nasa.gov/gcn/gcn3/26472.gcn3}

\bibitem[{Hess(1912)}]{hess}
Hess, V.~F. 1912, Physikalische Zeitschrift, 13, 1084–1091

\bibitem[{Hirata {et~al.}(1987)Hirata, Kajita, Koshiba, Nakahata, Oyama, Sato,
  Suzuki, Takita, Totsuka, Kifune, Suda, Takahashi, Tanimori, Miyano, Yamada,
  Beier, Feldscher, Kim, Mann, Newcomer, Van, Zhang, \&
  Cortez}]{PhysRevLett.58.1490}
Hirata, K., Kajita, T., Koshiba, M., {et~al.} 1987, Phys. Rev. Lett., 58, 1490,
  \dodoi{10.1103/PhysRevLett.58.1490}

\bibitem[{Hoskin(1999)}]{hoskin1999cambridge}
Hoskin, M. 1999, The Cambridge Concise History of Astronomy (Cambridge
  University Press).
\newblock \url{https://books.google.com/books?id=qDTSBgAAQBAJ}

\bibitem[{Jansky(1933)}]{radio}
Jansky, K.~G. 1933, Nature, 132, 66

\bibitem[{Keivani {et~al.}(2019)Keivani, Veske, Countryman, Bartos, Corley,
  Marka, \& Marka}]{Keivani:2019smf}
Keivani, A., Veske, D., Countryman, S., {et~al.} 2019, in {Proceedings of 36th
  International Cosmic Ray Conference - PoS(ICRC2019)}, Vol. 358, 930.
\newblock \doarXiv{1908.04996}

\bibitem[{Kimura {et~al.}(2017)Kimura, Murase, Mészáros, \&
  Kiuchi}]{Kimura_2017}
Kimura, S.~S., Murase, K., Mészáros, P., \& Kiuchi, K. 2017, The
  Astrophysical Journal Letters, 848, L4, \dodoi{10.3847/2041-8213/aa8d14}

\bibitem[{Magli(2016)}]{gobekli}
Magli, G. 2016, Nexus Netw J, 18, 337–346, \dodoi{10.1007/s00004-015-0277-1}

\bibitem[{Meegan {et~al.}(2009)Meegan, Lichti, Bhat, Bissaldi, Briggs,
  Connaughton, Diehl, Fishman, Greiner, Hoover, \& et~al.}]{Meegan_2009}
Meegan, C., Lichti, G., Bhat, P.~N., {et~al.} 2009, The Astrophysical Journal,
  702, 791–804, \dodoi{10.1088/0004-637x/702/1/791}

\bibitem[{{Neyman} \& {Pearson}(1933)}]{10.1098/rsta.1933.0009}
{Neyman}, J., \& {Pearson}, E.~S. 1933, Philosophical Transactions of the Royal
  Society of London Series A, 231, 289, \dodoi{10.1098/rsta.1933.0009}

\bibitem[{Opal {et~al.}(1974)Opal, Carruthers, Prinz, \& Meier}]{Opal702}
Opal, C.~B., Carruthers, G.~R., Prinz, D.~K., \& Meier, R.~R. 1974, Science,
  185, 702, \dodoi{10.1126/science.185.4152.702}

\bibitem[{{Penzias} \& {Wilson}(1965)}]{1965ApJ...142..419P}
{Penzias}, A.~A., \& {Wilson}, R.~W. 1965, The Astrophysical Journal, 142, 419,
  \dodoi{10.1086/148307}

\bibitem[{Rieke(2009)}]{ir}
Rieke, G. 2009, Exp Astron, 25, 125, \dodoi{/10.1007/s10686-009-9148-7}

\bibitem[{Sommers \& Westerhoff(2009)}]{Sommers_2009}
Sommers, P., \& Westerhoff, S. 2009, New Journal of Physics, 11, 055004,
  \dodoi{10.1088/1367-2630/11/5/055004}

\bibitem[{Urban(2016)}]{raven}
Urban, A.~L. 2016, PhD thesis, University of Wisconsin Milwaukee.
\newblock \url{dc.uwm.edu/etd/1218}

\bibitem[{Veske {et~al.}(2020)Veske, Márka, Bartos, \& Márka}]{Veske_2020}
Veske, D., Márka, Z., Bartos, I., \& Márka, S. 2020, Journal of Cosmology and
  Astroparticle Physics, 2020, 016–016, \dodoi{10.1088/1475-7516/2020/05/016}

\bibitem[{{Waxman} \& {Bahcall}(1997)}]{1997PhRvL..78.2292W}
{Waxman}, E., \& {Bahcall}, J. 1997, \prl, 78, 2292,
  \dodoi{10.1103/PhysRevLett.78.2292}

\end{thebibliography}
\bibliographystyle{aasjournal}

\end{document}